\documentclass[12pt,english,aps,preprint]{revtex4}
\usepackage[T1]{fontenc}
\usepackage[latin1]{inputenc}
\usepackage{graphicx}

\makeatletter

 \newenvironment{lyxlist}[1]
   {\begin{list}{}
     {\settowidth{\labelwidth}{#1}
      \setlength{\leftmargin}{\labelwidth}
      \addtolength{\leftmargin}{\labelsep}
      }}
   {\end{list}}

\def\Tr   {\mathop{\hbox{Tr}}}

\usepackage{babel}
\makeatother
\begin{document}
\begin{center}\textbf{\Large Simulating Full QCD with the Fixed Point
Action}\\
\end{center}{\Large \par}

\begin{center}\textbf{\textit{BGR collaboration}}\end{center}

\author{Anna Hasenfratz}

\email{anna@eotvos.colorado.edu}

\thanks{}

\affiliation{Department of Physics, University of Colorado, Boulder, CO-80304-0390}

\author{Peter Hasenfratz}

\email{hasenfra@itp.unibe.ch }

\thanks{}

\affiliation{Institute for Theoretical Physics, University of Bern, CH-3012, Bern,
Switzerland }

\author{Ferenc Niedermayer}

\email{niederma@itp.unibe.ch }

\thanks{On leave from HAS Res. Inst., E\"otv\"os University, Budapest}

\affiliation{Institute for Theoretical Physics, University of Bern, CH-3012, Bern,
Switzerland}

\begin{abstract}
Due to its complex structure the parametrized fixed point action can
not be simulated with the available local updating algorithms. We
constructed, coded, and tested an updating procedure with 2+1 light
flavors, where the targeted $s$ quark mass is at its physical value
while the $u$ and $d$ quarks should produce pions lighter than $300\,$MeV.
In the algorithm a partially global gauge update is followed by several
accept/reject steps, where parts of the determinant are switched on
gradually in the order of their expenses. The trial configuration
that is offered in the last, most expensive, stochastic accept/reject
step differs from the original configuration by a Metropolis + over-relaxation
gauge update over a sub-volume of $\sim(1.3\,\mathrm{fm})^{4}$. The
acceptance rate in this accept/reject step is $\sim0.4$. The code
is optimized on different architectures and is running on lattices
with $L_{s}\approx1.2\,\mathrm{fm}$ and $1.8\,\mathrm{fm}$ at a
resolution of $a\approx0.15\,\mathrm{fm}$.

\end{abstract}
\maketitle

\section{introduction}

There is a considerable freedom in formulating QCD on the lattice.
This freedom is reflected in the large number of actions tested and
used in the quenched approximation. There are no miracles: good scaling,
good chiral properties, theoretical safety and the expenses are in
balance. Short of an algorithmic breakthrough one can expect to see
in the future a plethora of full QCD simulations and results  obtained
with different formulations adapted to the physical problem as it
happened in the quenched approximation.

In this paper we discuss a full QCD algorithm for 2+1 light flavors
with the parametrized fixed point action \cite{Hasenfratz:2000xz,Niedermayer:2000yx,Jorg:2002nm,Hauswirth:2002eq}.
The lightest quark mass $m_{ud}$ which can be simulated is set only
by the small chiral symmetry breaking caused by the parametrization
error, the expenses of a full updating sweep are practically independent
of $m_{ud}$. 

Our updating procedure has no special problems in connecting different
topological sectors nor in suppressing the topological susceptibility.
  The algorithm is exact and the action certainly describes QCD in
the continuum limit. It is a partially global update where the pieces
of the determinant are switched on gradually in the order of their
expenses. The \textit{partially} global update implies that this is
a volume-squared algorithm which constraints the size of lattices
one can cope with in the simulations. Actually, the expense of a specific
section of the algorithm, the eigenvalue solver, increases even faster
with the volume, but in the present simulations this part is not dominating. 

In the ongoing runs the target spatial sizes are $L_{s}\approx1.2\,$fm
and $1.8\,$fm with a resolution $a\approx0.15\,$fm. The target pion
mass is below $300\,$MeV. On the $1.2\,$fm lattice we also ran with
the smallest $m_{ud}$ quark mass which can be treated with our Dirac
operator, basically with massless quarks. 

Actions which approximate the fixed point of a renormalization group
transformation have been tested in detail in $d=2$ and $d=4$. The
present form of the QCD fixed point action and the related codes are
the result of a long development to which many of our colleagues contributed
(see Ref. \cite{Gattringer:2003qx} and references therein). The algorithm
discussed here is optimized and running on three different platforms
(IBM SP4, PC Cluster, Hitachi SR8000). This paper is organized as
follows.  In Sect.\ref{sec:The-Action} we briefly describe the fixed
point action. Sect.\ref{sec:The-Stochastic-Update} summarizes the
Partial Global Stochastic Update and its improvements in a general
form while Sect.\ref{sec:The-2+1-Algorithm} describes how the update
is implemented in our simulation. In Sect.\ref{sec:Preliminary-Results}
we present some numerical results that illustrate the updating algorithm.

\section{The Action\label{sec:The-Action}}

\subsection{The parametrized fixed point action\label{sub:The-parametrized-fixed}}

The special QCD action which is the fixed point of a renormalization
group transformation has several desirable properties \cite{Hasenfratz:1993sp}.
It is a local solution of the Ginsparg-Wilson equation \cite{Ginsparg:1981bj},
\begin{equation}
\gamma_{5}D+D\gamma_{5}=D\gamma_{5}2RD,\label{G-W_eq}
\end{equation}
 with a non-trivial local matrix $R$ which lives on the hypercube
\cite{Hasenfratz:1997ft}. The quark mass is introduced as 
\begin{equation}
D(m)=D+m\left(\frac{1}{2R}-\frac{1}{2}D\right).\label{D(m)_deff}
\end{equation}
Eq.\ref{G-W_eq} guarantees that the Dirac operator is chirally invariant.
Since it is the fixed point of a renormalization group transformation,
it has no cut-off effects in the classical limit. The parametrized
version of this action is an approximation which has been carefully
tested in the classical limit and in quenched simulations \cite{Hasenfratz:2001qp,Hasenfratz:2002rp,Gattringer:2003qx,Hasenfratz:2004qk}.

The parametrized fixed point gauge action $S_{g}(U)$ \cite{Niedermayer:2000yx}
is a function of plaquette traces built from the original gauge links
$U$, and from smeared links. The smeared link contains staples in
an asymmetric way: the weights of staples which lie in, or orthogonal
to the plane of the plaquette are different. The gauge action is a
polynomial of smeared and unsmeared plaquette traces. The 5 non-linear
and 14 linear parameters are fitted to the fixed point action.

The Dirac operator \cite{Hasenfratz:2000xz,Hasenfratz:2001hr} is
constructed on smeared gauge configurations $V(U)$. This smearing
is local and contains links projected to SU(3). It is constructed
using renormalization group considerations \cite{DeGrand:1997ss}
and it reflects the discontinuous character of the chiral Dirac operator
when the topological charge changes \cite{Jorg:2002nm}. The Dirac
operator has fermion offsets on the hypercube only. In Dirac space
all the elements of the Clifford algebra enter. The structure of these
terms is restricted by the symmetries $C$, $P$, $\gamma_{5}$-hermiticity,
and cubic symmetry. The 82 free coefficients of this Dirac operator
are determined by a fit to the fixed point Dirac operator. 
We note that also for the Chirally Improved Dirac operator -- which has
a similarly complex structure -- a global update has been suggested
in Ref. \cite{Lang:2004ze}

\subsection{The 2+1 flavor action\label{sub:The-2+1-flavor}}

Our goal is to simulate a $2+1$ flavor system with quark masses close
to their physical values. As usual we integrate out the the fermionic
fields and write the action as\begin{eqnarray}
S & = & \beta S_{g}(U)+\bar{u}D(m_{ud})u+\bar{d}D(m_{ud})d+\bar{s}D(m_{s})s\label{S0}\\
 & \simeq & \beta S_{g}(U)-\ln{\rm {det}^{2}}{D(m_{ud})}-\ln\det{D(m_{s})}.\label{S1}\end{eqnarray}
The $\gamma_{5}$-hermiticity of the Dirac operator ensures that $\det{D(m)}$
is real, $\det{D(m)}=\det{D^{\dagger}(m)}.$ If $D$ were an exact
solution of Eq.\ref{G-W_eq} then $\det{D(m)}$ would be positive
for any $m>0$. Due to parametrization errors our Dirac operator has
no exact chiral symmetry and the requirement of $\det{D(m)}>0$ puts
a constraint on the quark mass $m$. For now we assume that this condition
is satisfied. 

One can rewrite the action in an explicitly Hermitian form 
\begin{equation}
S=\beta
S_{g}(U)-\ln{\det{(D^{\dagger}(m_{ud})D(m_{ud}))}}-\ln{\det{\left(\sqrt{D^{\dagger}(m_{s})}\sqrt{D(m_{s})}\right)}}\,.\label{action_2+1}
\end{equation}
 We want to emphasize that the appearance of the square root in Eq.\ref{action_2+1}
does not mean that we simulate a possibly non-local action. The action
we simulate is given by the manifestly local form in Eq.\ref{S0}.

\section{The Stochastic Update and its Improvements\label{sec:The-Stochastic-Update}}

The parametrized fixed point action contains many gauge paths and
SU(3) projections. It is too complicated, if not impossible, to simulate
with algorithms that would require the derivative of the action with
respect of the gauge fields. For that reason we have adapted an updating
method that requires only a stochastic estimate of the action at each
step. The (Partial) Global Stochastic Update was developed in \cite{Hasenfratz:2002jn},
based on an old suggestion \cite{Grady:1985fs,Creutz:Book}, to simulate
smeared link staggered actions. It was developed further in \cite{Hasenfratz:2002ym,Alexandru:2002jr,Knechtli:2003yt}.
Stochastic or {}``noisy'' updating algorithms have been used in
different context by many other groups \cite{Kennedy:1985pg,Kennedy:1988yy,Joo:2001bz,Montvay:2005tj}.
Here we take over the main points of the Global Stochastic Update
but add several improvements to create an efficient updating algorithm.
In the next section we will summarize the general ideas of the update,
then discuss the specific improvements we have implemented.

\subsection{The Partial-Global Stochastic Update}

In order to simplify the notation in this section we consider an action
with the generic form 
\begin{equation}
S=\beta S_{g}(U)-\ln{\det{A^{\dagger}A}}.\label{action_gen1}
\end{equation}
 Here $A^{\dagger}A$ describes 1 or 2 flavors of massive fermions
as discussed in Sect.\ref{sub:The-2+1-flavor}. A Global Update proceeds
in two steps:

\begin{lyxlist}{00.00.0000}
\item [A:]Update (a part of) the configuration $U\to U'$ with the gauge
action $\beta S_{g}$ using Metropolis, over-relaxation or other  updates.
\end{lyxlist}
One could accept or reject (A/R) the proposed $U'$ configuration
with probability
\begin{equation}
P_{{\rm {acc}}}={\rm {min}\left(1,\frac{\det A'^{\dagger}A'}{\det
      A^{\dagger}A}\right).}
\end{equation}
This procedure clearly satisfies the detailed balance condition but
requires the evaluation of the fermionic determinant. This lengthy
calculation can be replaced by a stochastic estimator as follows.
We write the determinant ratio as a stochastic integral \begin{eqnarray}
\frac{\det A'^{\dagger}A'}{\det A^{\dagger}A} & = & \det{(\Omega{}^{\dagger}\Omega)^{-1}}\nonumber \\
 & \simeq & \int D[\eta^{\dagger}\eta]e^{-\eta^{\dagger}\eta}e^{-(\eta^{\dagger}\Omega^{\dagger}\Omega\eta-\eta^{\dagger}\eta)}\end{eqnarray}
 where we introduced the notation 
\begin{equation}
\Omega=A'^{-1}A\,.
\end{equation}

\begin{lyxlist}{00.00.0000}
\item [B:]For the stochastic accept/reject step we first create a Gaussian
random vector $\eta$ with $P(\eta)\propto\exp(-\eta^{\dagger}\eta)$.
Now the new configuration is accepted with the probability
\begin{equation}
P_{{\rm {acc}}}^{{\rm {stoch}}}=\min{(1,e^{-\Delta S_{f}})},\label{P_stoch}
\end{equation}
 where
\begin{equation}
\Delta S_{f}=\eta^{\dagger}(\Omega^{\dagger}\Omega-1)\eta.\label{delta_Sf}
\end{equation}
Eq.\ref{delta_Sf} defines the stochastic estimator $\Delta S_{f}$,
the change of the fermionic action with fixed $\eta$ \cite{Hasenfratz:2002jn}. 
\end{lyxlist}
The stochastic update satisfies the detailed balance condition \cite{Grady:1985fs,Creutz:Book,Alexandru:2002jr},
and repeating steps A-B creates a sequence of gauge configurations
with the proper probability distribution. 

For the 2+1 flavor system of Eq.\ref{action_2+1} the stochastic estimator
contains two terms 
\begin{equation}
\Delta
S_{f}=\eta_{u}^{\dagger}(\Omega_{ud}^{\dagger}\Omega_{ud}-1)\eta_{u}+\eta_{s}^{\dagger}(\Omega_{s}^{\dagger}\Omega_{s}-1)\eta_{s}
\end{equation}
 with $\Omega_{ud}=D'\,^{-1}(m_{ud})D(m_{ud})$ and $\Omega_{s}=\sqrt{D'^{-1}(m_{s})}\sqrt{D(m_{s})}$. 

The main ideas of the stochastic Monte Carlo update has been known
for a long time but it has not been used in numerical simulations
because of technical difficulties. In its original form the acceptance
rate in Eq.\ref{P_stoch} is infinitesimally small unless the configurations
$U$ and $U'$ are nearly identical. The root of this problem is two-fold.

On one hand, if the typical values of $|\log\det(D'/D)|$ are significantly
larger than 1 the acceptance rate will be very small, even if the
determinant ratio was calculated deterministically. On the other hand,
an additional suppression of the acceptance rate occurs due to the
stochastic evaluation in the A/R step. To illustrate this consider
the case when $\det D'/\det D=1$. Take a simple model for this situation
with a Gaussian random variable $P(x)\propto\exp(-(x-x_{0})^{2}/2\sigma^{2})$.
The relation $\langle\mathrm{{e}^{-x}\rangle=1}$ implies $x_{0}=\sigma^{2}/2$
hence for large $\sigma$ the acceptance rate is extremely small,
$\sim\mathrm{e^{-\sigma^{2}/2}\ll1}$.

Note that the standard deviation of $\exp(-\eta^{\dagger}(\Omega^{\dagger}\Omega-1)\eta)$
is infinite if any of the eigenvalues of $\Omega=A'^{-1}A$ is smaller
than $1/2$ \cite{Alexandru:2002jr}. However, the extremely small
acceptance rate occurs much before this bound is reached. In the following
we discuss four improvement steps which are essential to get an algorithm
with a good acceptance rate.

\subsection{Improvements of the Global Stochastic Update}

In this section we discuss the different improvements we implemented
to increase the effectiveness of the stochastic updating. In the \textit{reduction}
technique the UV part of the determinant is separated, its value is
calculated non-stochastically and taken into account more frequently
by intermediate A/R steps. This increases the acceptance rate in the
stochastic A/R step significantly by reducing both problems mentioned
above. The \textit{subtraction} technique separates the IR modes by
calculating the eigenvalues and eigenvectors of the first few low
eigenmodes of the Dirac operator. It acts analogously for the IR modes
as the reduction for the UV part. The last two techniques, the relative
gauge fixing and the determinant breakup, are applied in the last,
stochastic A/R step, and are aimed at reducing the fluctuations. The
\emph{relative gauge fixing} brings the configuration $U'$ as close
to $U$ as possible. The \textit{determinant} \textit{breakup} rewrites
the Dirac operator as product of operators. The stochastic estimator
becomes a sum of independent terms and its fluctuation is reduced.

The Global Stochastic Monte Carlo Update would not be effective without
these improvements. On large lattices even with the improvements one
can update only a part of the configuration before evaluating the
stochastic estimator making the algorithm to scale with the square
of the volume. Nevertheless on moderate volumes we found the algorithm
effective, allowing the dynamical simulation of light, even nearly
massless, quarks with an action where the chiral breaking and lattice
artifacts are small.

\subsubsection{The Reduction:}

The stochastic change of the fermionic action can be written as $\Delta S_{f}=\sum_{i}(\omega_{i}-1)\eta_{i}^{\dagger}\eta_{i}$
where $\omega_{i}$ are the real eigenvalues of the operator $\Omega^{\dagger}\Omega$
and $\eta_{i}^{\dagger}\eta_{i}=O(1)$. While the eigenvalues of the
Dirac operator are restricted to a compact region, $\omega_{i}$ can
vary between $\sim m$ to $\sim1/m$, though most of the eigenvalues
correspond the the UV modes are $O(1)$. These UV modes contribute
little to $\Delta S_{f}$ individually, but there are so many of them
that they dominate the fluctuations. To reduce the fluctuations we
transform the Dirac operator $D\to D_{r}$ such that the UV modes
of $D_{r}$ are condensed and thus the corresponding eigenmodes of
$\Omega$ are closer to unity. We choose $D_{r}$ such that the change
in the determinant $\det(D/D_{r})$ is calculable analytically (non-stochastically). 

\begin{figure}
\includegraphics[%
  width=12cm]{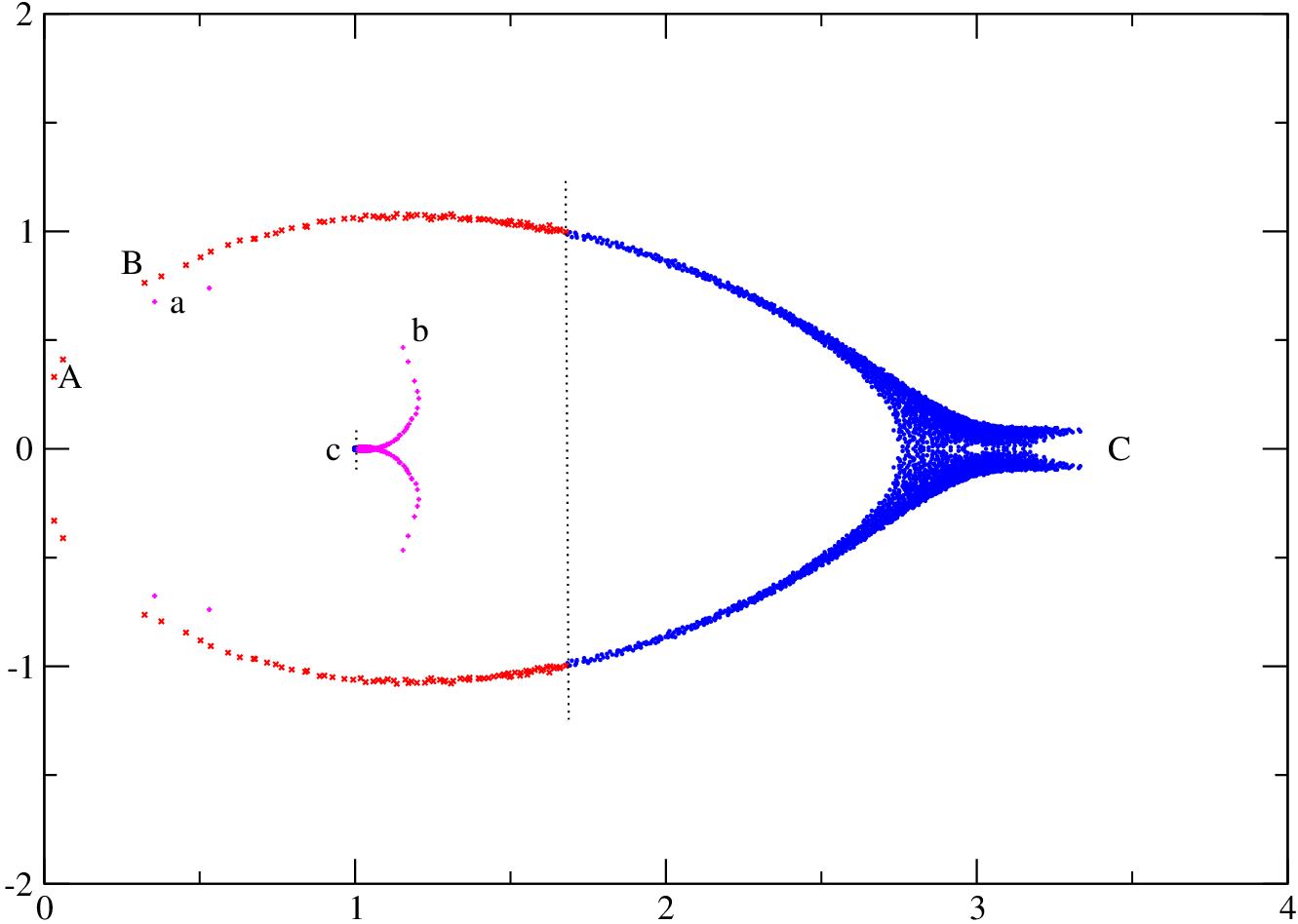}

\caption{The eigenvalue spectrum of the Dirac operator on a single $4^{4}$
gauge configuration. The points of the larger {}``Batman'' like
figure correspond to the original Dirac operator while the points
of the smaller {}``wing-shape'' structure in the center represent
the corresponding eigenvalues of the reduced Dirac operator $D_{r}/s$.
Sections marked by $A$, $B$ and $C$ on the original spectrum are
mapped to sections $a$, $b$ and $c$ after reduction.\label{Fig:spect_1}}
\end{figure}
The effect of the reduction on the eigenvalue spectrum of $D$ is
illustrated in Figure \ref{Fig:spect_1}. The spectrum was calculated
on a single $4^{4}$ quenched gauge configuration $U$ with $a\approx0.15\,$fm
in \cite{Jorg:2002nm}. The larger {}``Batman'' like structure corresponds
to the spectrum of the original Dirac operator. The {}``Batman ears''
are mainly the consequence of the non-trivial $R$ operator and are
strongly reduced if one considers the spectrum of $DR$ which is close
to a circle. An overwhelming part of the eigenvalues are around the
$UV$ point $(s,0)$ in the complex plane where $s\approx2.8$. The
points to the right of the long dotted line represent 95\% of the
eigenvalues. With the reduction we attempt to move the eigenvalues
of $D/s$ close to $1$. 
Following Refs. \cite{deForcrand:1998sv,Hasenbusch:2001ne,Hasenfratz:2002jn}
we define a reduced Dirac operator 
\begin{equation}
D_{r}=D\, e^{-\sum_{i=1}^{n}c_{i}(D/s-1)^{i}}.\label{D_reduced}
\end{equation}
 Choosing the coefficients $c_{i}=(-1)^{i+1}/i$ the reduced operator
is $D_{r}/s=1+O((D/s-1)^{n+1})$. The ratio of the determinants $D_{r}$
and $D$ can be expressed in terms of traces of $D$\begin{eqnarray}
\det D_{r} & = & \det D\, e^{-\sum_{i=1}^{n}c_{i}Tr(D/s-1)^{i}}\nonumber \\
 & = & \det D\, e^{-\sum_{i=1}^{n}\alpha_{i}TrD^{i}}\,,\label{S3}\end{eqnarray}
 and can be calculated non-stochastically by evaluating $\Tr D^{k}$,
$k=1,n$. The computing time and the complexity of the code to do
the trace calculations increase rapidly as $n$ increases. With our
Dirac operator we decided to stop at $n=4$ in the reduction. Working
with $D_{r}$ is not much different from $D$. Both in multiplication
and inversion the exponential term in Eq.\ref{D_reduced} can be approximated
by a relatively low order polynomial. 

In Figure \ref{Fig:spect_1} the smaller wing-shape object in the
center corresponds to the eigenvalues of $D_{r}/s$ on the same gauge
configuration as before. Sections marked as $A$, $B$ and $C$ on
the original spectrum are mapped to sections $a$, $b$ and $c$ after
reduction, illustrating how the eigenvalues are transformed. The origin
is a stationary point. The main effect of the reduction is condensing
the UV modes. The points to the left of the short dotted line again
represent 95\% of the eigenvalues. This region is not very obvious
in the figure since all its points are contained in a circle around
$(1,0)$ with radius of about 0.01 (with the overwhelming part of
the eigenvalues being much closer), showing the strength of the reduction.
Accordingly, the $UV$ fluctuations, that is the contribution of the
UV modes to the stochastic estimator $\Delta S_{f}$, is greatly reduced. 

At this point the action Eq.\ref{action_2+1} can be written as \begin{eqnarray}
S & = & \beta S_{g}(U)+S_{UV}\nonumber \\
 & - & \ln{\det{D_{r}^{\dagger}(m_{ud})D_{r}(m_{ud})}}-\ln{\det{D_{r}(m_{s})}}\label{S_2+1_UV}\end{eqnarray}
 where 
\begin{equation}
S_{UV}=-2\sum_{i=1}^{n}\alpha_{i}\Tr{D^{i}(m_{ud})}-\sum_{i=1}^{n}\alpha_{i}\Tr{D^{i}(m_{s})}\label{S_UV}
\end{equation}
 carries most of the UV part of the determinant.

\subsubsection{The Subtraction}

The reduction of the Dirac operator as discussed in the previous section
effects mainly the non-physical UV part of the determinant. The subtraction
that we introduce here deals with the low lying IR eigenvalues of
the Dirac operator. The small eigenvalues of $D'$ can create large
$\omega_{i}$ eigenvalues of $\Omega=D'\,^{-1}D$. Besides suppressing
such configurations in the (full QCD) equilibrium configurations,
their presence produces large fluctuations in the stochastic estimator,
reducing therefore the acceptance rate in the stochastic A/R step.
By calculating some low lying eigenvalues (and the corresponding eigenvectors)
one can take into account their contribution more frequently and deterministically,
so they do not participate in the stochastic A/R step. 

Denote the right and left eigenvectors of the Dirac operator by 
\begin{equation}
Dv_{\lambda}=\lambda v_{\lambda}\,,\qquad w_{\lambda}^{\dagger}D=\lambda
w_{\lambda}^{\dagger}\,.
\end{equation}
 The eigenvectors $w_{\lambda}^{\dagger}$ can be chosen to fulfill
the normalization condition 
\begin{equation}
w_{\lambda}^{\dagger}v_{\lambda'}=\delta_{\lambda\lambda'}\,.\label{norm}
\end{equation}
In terms of the (non-hermitian) projector operators 
\begin{equation}
P_{\lambda}=v_{\lambda}w_{\lambda}^{\dagger}\,
\end{equation}
 which, due to Eq.\ref{norm}, satisfy the relation $P_{\lambda}^{2}=P_{\lambda}$,
the Dirac operator can be written as 
\begin{equation}
D=\sum\lambda P_{\lambda}\,.
\end{equation}
The subtracted Dirac operator is defined by replacing a set of the
lowest eigenvalues by the constant $s$ 
\begin{equation}
D_{s}=D+\sum_{{\rm {low}}}(s-\lambda)P_{\lambda}\,.\label{D_subtracted}
\end{equation}
 We assume that we subtract the complex conjugate pairs together.
For an arbitrary analytic function $f(D)$ the subtraction gives 
\begin{equation}
f(D_{s})=f(D)+\sum_{{\rm
    {low}}}(f(s)-f(\lambda))P_{\lambda}.\label{Subtraction_def}
\end{equation}
Subtracting the reduced Dirac operator of Eq. \ref{D_reduced} gives
then \[
D_{rs}=D\mathrm{e}^{-\sum_{i}c_{i}(D/s-1)^{i}}+\sum_{{\rm {low}}}\left(s-\lambda\mathrm{e}^{-\sum_{i}c_{i}(\lambda/s-1)^{i}}\right)P_{\lambda}\,.\]
The small eigenvalues of the subtracted, reduced $D_{rs}$ operator
are replaced by $s$ while its UV part is condensed near $s$. The
stochastic estimator of $D_{rs}$ has reduced fluctuations and reduced
absolute value as well. 

The ratio of the determinants of $D_{rs}$ and $D_{r}$ can be calculated
analytically using the relation 
\begin{equation}
\det D_{r}=\det D_{rs}\prod_{{\rm
    {low}}}\frac{\lambda}{s}e^{-\sum_{i}c_{i}(\lambda/s-1)^{i}}\,.
\end{equation}

The inversion of $D$ using that of $D_{s}$ is given by 
\begin{equation}
D^{-1}=\left[1+\sum_{{\rm
      {low}}}\left(\frac{s}{\lambda}-1\right)P_{\lambda}\right]D_{s}^{-1}\,.
\end{equation}
 The smallest eigenvalues of $D$ are replaced by the constant $s$
in $D_{s}$ therefore the conjugate gradient method converges faster
for $D_{s}$. 

At this point the action of Eqs.\ref{action_2+1} and \ref{S_2+1_UV}
can be written as \begin{eqnarray}
S & = & \beta S_{g}(U)+S_{UV}+S_{IR}\nonumber \\
 & - & \ln{\det{D_{rs}^{\dagger}(m_{ud})D_{rs}(m_{ud})}}-\ln{\det{D_{rs}(m_{s})}}\label{S_2+1_UV_IR}\end{eqnarray}
where\begin{eqnarray}
S_{IR} & = & 2\sum_{{\rm {low}}}\left(-\ln\frac{|\lambda_{ud}|}{s}+\sum_{i}c_{i}\left(\frac{\lambda_{ud}}{s}-1\right)^{i}\right)\nonumber \\
 & + & \sum_{{\rm {low}}}\left(-\ln\frac{|\lambda_{s}|}{s}+\sum_{i}c_{i}\left(\frac{\lambda_{s}}{s}-1\right)^{i}\right)\,.\label{S_IR}\end{eqnarray}

\subsubsection{The Relative Gauge Fixing\label{sub:The-Relative-Gauge}}

Even if $U'$ is a gauge transform of $U$, hence the determinant
ratio is exactly 1, the eigenvalues of $\Omega=D(U')^{-1}D(U)$ are
in general different from 1 (only their product is 1). As a consequence
the stochastic estimator in the A/R step can have large fluctuations,
greatly reducing the acceptance rate\cite{Knechtli:2003yt}. These
fluctuations can be reduced significantly by gauge transforming $U'$
so as to maximize $\mathrm{\sum_{x,\mu}{Re}\mathrm{{Tr}(U'_{x\mu}U_{x\mu}^{\dagger})}}$,
i.e. by bringing $U'$ as close to $U$ as possible. (We have also
tried to fix the gauge in both $U$ and $U'$ by some given {}``absolute''
gauge fixing condition, but the method discussed above was more efficient.)

\subsubsection{The Determinant Breakup\label{sub:The-Determinant-Breakup}}

The reduction, subtraction, and relative gauge fixing result in significant
improvement of the stochastic estimator. Further improvements can
be achieved by writing the Dirac operator as the product of $l$ terms
\begin{equation}
A=A_{1}\times A_{2}\times....\times A_{l}\,.\label{Breakup}
\end{equation}
 The corresponding stochastic estimator is the sum of $l$ terms 
\begin{equation}
\Delta
S_{f}=\sum_{i=1}^{l}\eta_{i}^{\dagger}(\Omega_{i}^{\dagger}\Omega_{i}-1)\eta_{i}\label{Stoch_est_breakup}
\end{equation}
 with $l$ stochastic $\eta$ vectors and $\Omega_{i}=A'_{i}\,^{-1}A_{i}$.
If the eigenspectrum of the individual $A_{i}$ operators is closer
to a constant the fluctuation of the stochastic estimator is reduced.
In Refs. \cite{Alexandru:2002jr,Alexandru:2002sw}, following a suggestion
in Ref. \cite{Hasenbusch:1998yb}, the terms in Eq.\ref{Breakup}
were chosen to be identical, $A_{i}=A^{1/l}$. While this choice does
reduce the fluctuations, it creates $l$ equally singular terms and
requires the calculation of the $l$th root for each of them. We found
it is more effective to generalize the mass shifting method of \cite{Hasenbusch:2001ne}
and write the Dirac operator as 
\begin{equation}
\frac{1}{s}A=\frac{A(\mu_{0})}{A(\mu_{1})}\times\frac{A(\mu_{1})}{A(\mu_{2})}\times...\times\frac{A(\mu_{l-1})}{A(\mu_{l})}\times
A(\mu_{l})\,,
\end{equation}
 where $A(\mu)=\frac{1}{s+\mu}(A+\mu)$ and $\mu_{0}=0$. The mass
shift values $\mu_{i}$ are chosen such that each term in $\Delta S_{f}$
contributes approximately equally. 

The first term in Eq.\ref{Stoch_est_breakup} is the easiest to analyze.
At lowest order in $\mu_{1}$
\begin{equation}
\eta_{1}^{\dagger}(\Omega_{1}^{\dagger}\Omega_{1}-1)\eta_{1}=-\eta_{1}^{\dagger}\left(\frac{\mu_{1}}{A}+\frac{\mu_{1}}{A^{\dagger}}-\frac{\mu_{1}}{A'}-\frac{\mu_{1}}{A'\,^{\dagger}}\right)\eta+O(\mu_{1}^{2})\,.\label{Stoch_1st_term}
\end{equation}
 Unless the operators $A$ and $A'$ are close to each other, $\mu_{1}/A$
has to be small to control the stochastic fluctuations. Consequently
$\mu_{1}$ has to be much smaller than the smallest eigenvalue of
$A$. Later terms allow larger change in the shift masses $\mu_{i}$.
The last mass of the series, $\mu_{l}$, is chosen such that the stochastic
estimator of the single operator $A(\mu_{l})$ is comparable to the
previous terms. It is interesting to note that the leading term in
Eq.\ref{Stoch_1st_term} vanishes for a Ginsparg-Wilson Dirac operator
with $R={\rm {const}}$. However it is not zero in our case.

In practice we combine the reduction, subtraction, relative gauge
fixing and the determinant breakup. Only the last two terms of Eq.\ref{S_2+1_UV_IR}
are treated stochastically. For the degenerate $u$ and $d$ quarks
we can have \begin{eqnarray}
A & = & D_{rs}(m_{ud})\,,\nonumber \\
A(\mu) & = & \left(\frac{D_{r}(m_{ud})+\mu}{s+\mu}\right)_{s}\,,\end{eqnarray}
 where the subtraction is defined as in Eq.\ref{Subtraction_def}.
Each term in the stochastic estimator requires two multiplications
and two inversions by $A(\mu)$. For the inversion a standard conjugate
gradient or its variant can be used. 

For the $s$ quark the situation is slightly more complicated. The
$A$ operator contains a square root operation\begin{eqnarray}
A & = & \sqrt{D_{rs}(m_{s})}\;,\nonumber \\
A(\mu) & = & \sqrt{\left(\frac{D_{r}(m_{s})+\mu}{s+\mu}\right)_{s}}\;.\\
\nonumber \end{eqnarray}
 Again, each term in the stochastic estimator requires two multiplications
and two inversions by $A(\mu)$. For both of these we approximate
the square root operator by a polynomial series. Polynomials have
been used to approximate both positive and negative roots of Dirac
operators before \cite{Montvay:1997vh,Alexandru:2002sw,Alexandru:2002jr}.
Our situation is different because the operator $D_{r}(m_{s})$ is
complex. 
The case of optimal polynomials for a complex spectrum has been studied
in \cite{Borici:1995am}.
Fortunately the strange quark mass is sufficiently heavy
and a Taylor expansion in $(D(m)/s-1)$ works well.

\section{The Algorithm with 2+1 Flavors\label{sec:The-2+1-Algorithm} }

We describe now the algorithm which has been coded, tested and optimized
for different platforms.The algorithm starts with a partially global
gauge update which is followed by several accept/reject steps, where
parts of the determinant are switched on gradually in the order of
their expenses. It is convenient to rewrite the action of Eq.\ref{S_2+1_UV_IR}
in a different form \begin{eqnarray}
S & = & (\beta+\delta\beta)S_{g}(U)\nonumber \\
 & + & [S_{UV}^{g}-\delta\beta S_{g}(U)]\nonumber \\
 & + & [S_{UV}-S_{UV}^{g}+S_{IR}^{{\rm {appr}}}]\\
 & + & [S_{IR}-S_{IR}^{{\rm {appr}}}-\ln{\det{D_{rs}^{\dagger}(m_{ud})D_{sr}(m_{ud})}}-\ln{\det{D_{rs}(m_{s})}}].\nonumber \end{eqnarray}
 The meaning of the different terms will be explained in the rest
of this section.

\subsection{Gauge update\label{sub:Gauge-update}}

The gauge update is a standard Metropolis/over-relaxation local update
with the fixed point gauge action at coupling $\beta_{{\rm {eff}}}=\beta+\delta\beta$.
Here $\delta\beta$ (added at this point and subtracted later) approximates
the shift of the gauge coupling due to the determinant and helps to
generate configurations with lattice spacing $a$ that is close to
the target value already at this step. $4n_{p}$ gauge links, originating
from $n_{p}$ consecutive lattice sites, are updated with Metropolis
and then the same links with over-relaxation in a reversible sequence.
This combination of updates we shall call a 'double update' in the
following.

In the first test runs our target lattice spacing is $a\approx0.15$
fm and $n_{p}$ is 128 and 144 on the $8^{3}\times24$ and $12^{3}\times24$
lattices, respectively. In order to get the resolution $a$ close
to the target value we had to tune the coupling $\beta$ repeatedly.
The figures in this paper refer to the choice $\beta=3.15$.

\subsection{The 1st accept/reject step\label{sub:The-1st-accept/reject}}

The gauge configuration created as discussed above is accepted/rejected
(A/R) with the action $S_{UV}^{g}-\delta\beta S_{g}(U)$, where $S_{UV}^{g}$
is a good gauge approximation to the reduction contribution $S_{UV}$
of Eq.\ref{S_UV}. The function $S_{UV}^{g}$ is represented by different
gauge loops with fitted coefficients on the smeared configuration
$V(U)$. This smearing is the same which was used in the parametrized
Dirac operator (Sect.\ref{sub:The-parametrized-fixed}). Calculating
$S_{UV}^{g}$ is fast and can be done without building up the Dirac
operator. The deviation between $S_{UV}^{g}$ and the exact reduction
$S_{UV}$ will be corrected in the 2nd accept/reject step below. The
parameter $\delta\beta$ is chosen to maximize the acceptance rate
in this step. %
\begin{figure}
\includegraphics[%
  width=10cm]{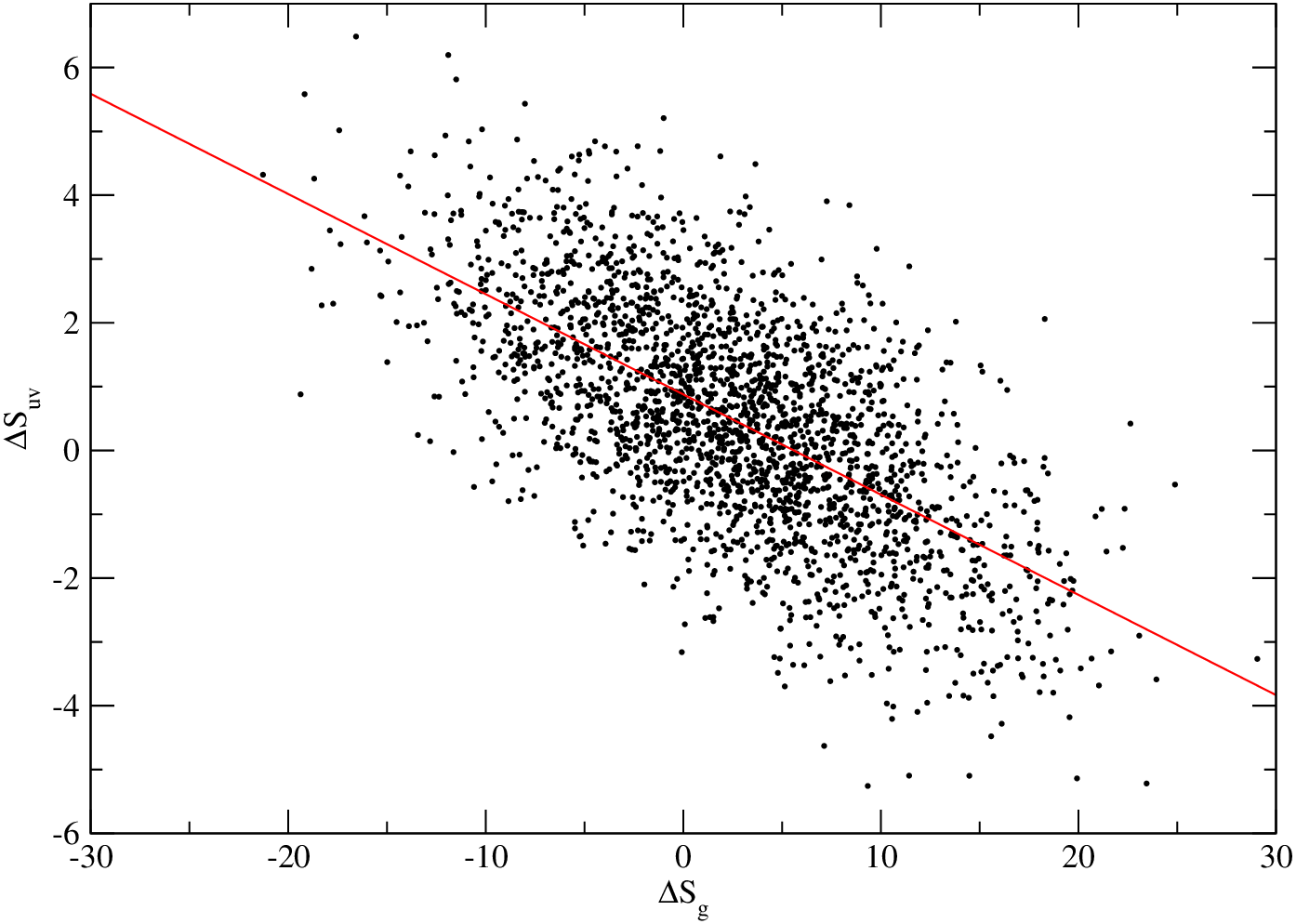}

\caption{The correlation between $\Delta S_{UV}$ and $\Delta S_{g}$. The
slope predicts the optimal value of $\delta\beta\approx0.15$. \label{Fig:S_UV-fit}}
\end{figure}
Figure \ref{Fig:S_UV-fit} shows the correlation between $\Delta S_{g}$,
the change of the gauge action, and $\Delta S_{UV}$, the change of
the contribution from the reduction, for a set of configuration pairs
$\{ U,U'\}.$ From the slope $\delta\beta=-0.15$ seems to be a reasonable
choice. It is interesting to note that for our action the introduction
of the determinant increases the gauge coupling. In the usual Wilson
and staggered fermion simulations this shift is larger and in the
opposite direction. Since the reduction contribution is large for
distant configurations and the $-\delta\beta S_{g}$ term cancels
it only approximately, we have to keep the number of updated links
$4n_{p}$ in the gauge update modest in order to get a good acceptance
rate in this 1st accept/reject. The combination of steps in Sects
\ref{sub:Gauge-update} and \ref{sub:The-1st-accept/reject} is repeated
$N_{1}$ times.

With the $n_{p}$ value quoted before, the 1st acceptance rate is
above 0.5. In the running simulations $N_{1}=28$, i.e. $4\times n\times N_{1}\approx15\mathrm{k}$
links are double updated before the 2nd accept/reject step.

\subsection{The 2nd accept/reject step\label{sub:The-2nd-accept/reject}}

\begin{figure}
\includegraphics[%
  width=10cm]{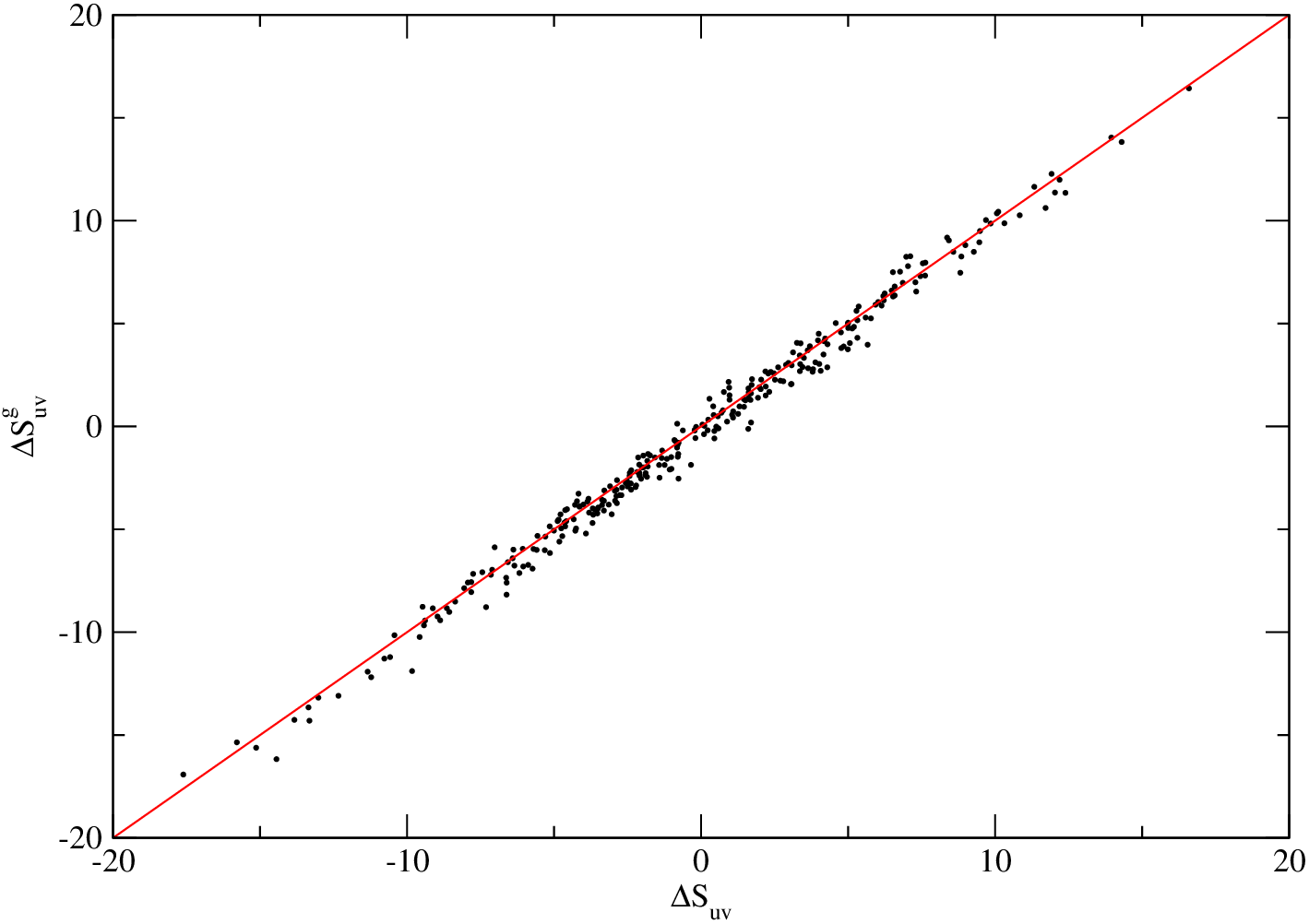}

\caption{The approximation of $\Delta S_{UV}$ in terms of gauge loops, $\Delta S_{UV}^{g}$.
This error is corrected in the 2nd A/R step. \label{Fig:dS_UV_g}}
\end{figure}

The cycle of repeated steps in Sects \ref{sub:Gauge-update} and \ref{sub:The-1st-accept/reject}
is followed by a 2nd accept/reject decision. In this step the Dirac
operator is built on the $U'$ competitor configuration, the traces
are calculated for the exact reduction, and a certain number of the
lowest eigenvalues and eigenvectors are determined. The proposed configuration
is accepted/rejected with the action $(S_{UV}-S_{UV}^{g})+S_{IR}^{{\rm {appr}}}$.
The first term corrects the small error we made in the 1st A/R step
in approximating the traces in $S_{UV}$ with gauge loops. This error
is typically small as shown in Figure \ref{Fig:dS_UV_g}, where the
change $\Delta S_{UV}^{g}$ calculated in the gauge approximation
is shown as the function of its exact value. (The action differences
plotted in this figure are taken between configurations which are
offered to the 3rd A/R step, as a result of several 2nd A/R steps.
These large values of O(10) would cause a very small acceptance rate
in the 3rd step, had we not taken into account this contribution more
frequently in the 2nd step.)%

The last term $S_{IR}^{{\rm {appr}}}$ is an approximation to the
contribution of the low lying eigenvalues to the determinant $S_{IR}$
in Eq.\ref{S_IR}. $n_{{\rm {ev}}}$ eigenvalues and the corresponding
$S_{IR}$ are calculated for $m_{ud}$ using an Arnoldi eigenvalue
finder. The eigenvalues for $m_{s}$ are determined from these, using
leading order perturbation theory. (Due to the presence of R in the
Ginsparg-Wilson relation, Eq.\ref{G-W_eq}, the reaction of the eigenvalues
on changing the quark mass is not a simple shift.) This approximation
is very good, the error is typically $O(10^{-4})$ . The combination
of the steps in Sects.\ref{sub:Gauge-update}, \ref{sub:The-1st-accept/reject}
and \ref{sub:The-2nd-accept/reject} is repeated $N_{2}$ times.

At present we calculate $n_{{\rm {ev}}}=48$ eigenvalues, which include
all the eigenvalues with absolute value below $\approx0.40$ and $\approx0.19$
on the $8^{3}\times24$ and $12^{3}\times24$ lattices, respectively.
The reduction shifts these values further to around $\approx1.0$
and $\approx0.5$. We use an Arnoldi routine from the publicly available
PARPACK package. The application is far from optimal in our case.
Even though the eigenvalues and eigenvectors are calculated on very
similar configurations and change little from step to step, the routine
cannot use this fact and calculates each eigenvalue set basically
independently. The internal calculations of the package are also expensive,
the multiplication of a vector by the Dirac operator does not dominate
it. These problems limit the number of times the 2nd A/R step is repeated.
We can afford $N_{2}=6$ repetitions and the acceptance rate of the
2nd A/R step is around 0.65. Overall about $\approx90\mathrm{k}$
links are double updated before the 3rd A/R step.

\subsection{The 3rd accept/reject step\label{sub:The-3rd-accept/reject}}

The cycle described above is followed by a final, stochastic accept/reject
step with the action $S_{IR}-S_{IR}^{{\rm {appr}}}-\ln{\det{D_{rs}^{\dagger}(m_{ud})D_{rs}(m_{ud})}}-\ln{\det{D_{rs}(m_{s})}}$.
The first part corrects the small error we made in calculating the
contribution of the low lying eigenvalues of $D(m_{s})$ to the determinant
in the 2nd A/R step. For that we determine the low lying spectrum
of $D(m_{s})$ on the competitor configuration $U'$ which is first
relative-gauge-fixed with respect to $U$. The second term gives the
stochastic estimator of the subtracted, reduced, 2+1 flavor determinant.
For the light quarks we break up the determinant into $76$ terms.
The lowest eigenvalue of the reduced subtracted Dirac operator, $D_{rs}/s$
is around one and the first few mass shifts have to be much smaller
than that (see Sect.\ref{sub:The-Determinant-Breakup}). We chose
$\Delta\mu_{i}=\mu_{i+1}-\mu_{i}=0.01$ for $i=1-20$. The later $\Delta\mu$
values are considerably larger. Due to the subtraction of the low
lying eigenmodes the conjugate gradient iteration converges relatively
fast, in about 70 steps at the lowest mass shifts and in 5-10 steps
at the largest, each step requiring two $D\times v$ Dirac operator
multiplications. The exponential term for the reduction requires about
20 $D\times v$ multiplications. For $m_{s}$ the determinant breakup
has 38 terms and the smallest shift is $\Delta\mu=0.02$. The square
root and its inverse of the reduced, subtracted Dirac operator is
approximated by their Taylor series in $(D/s-1)$. In the smaller
mass shift region we use 250-300 order polynomials, for the larger
mass shift values this reduces to order 30-40. We expect that this
section of the code could be significantly improved.

With this closing accept/reject step the algorithm becomes exact.
The steps in Sects. \ref{sub:Gauge-update}, \ref{sub:The-1st-accept/reject},
\ref{sub:The-2nd-accept/reject}, and \ref{sub:The-3rd-accept/reject}
are repeated and the accepted configurations that went through all
three filters form a Markov chain corresponding the parametrized fixed
point action. The acceptance rate of the 3rd A/R step is approximately
0.4.

\subsection{Optimization and performance }

The code is optimized on three different platforms (IBM SP4, PC cluster
and Hitachi SR8000). The dominating numerical step is the multiplication
of a vector by the Dirac operator, $D\times v$, which can be effectively
parallelized on all the platforms. The effectiveness of the internal
manipulations of the PARPACK package, however, is very sensitive to
the architecture. It would be very preferable to replace this part
of the code by a QCD specialized piece. 

The stochastic estimator (in the 3rd accept/reject) requires $\approx21\mathrm{k}$
and $\approx26\mathrm{k}$ $D\times v$ multiplications on the $8^{3}\times24$
and $12^{3}\times24$ lattices, respectively. To calculate the first
48 eigenvalues/eigenvectors of the Dirac operator requires $\approx1\mathrm{k}$
and $\approx2\mathrm{k}$ $D\times v$ multiplications on the smaller
and larger lattices, respectively. 

At certain stages of the calculation the processors are divided in
two groups and work on the gauge and Dirac part of the code independently.
They are joined, however, to calculate the stochastic estimator together.

\section{Preliminary Results\label{sec:Preliminary-Results}}

In our first set of test runs, starting from two different $a\approx0.15\,$fm
quenched configurations, we generated about 400 + 600 $8^{3}\times24$
configurations, each separated by a full cycle of updates and A/R
steps as described in Sects.\ref{sub:Gauge-update}-\ref{sub:The-3rd-accept/reject}.
We chose the run parameters, based on earlier quenched runs, as $\beta=3.15,$
$\delta\beta=-0.15,$ $m_{ud}=0.017$ and $m_{s}=0.095.$ %
\begin{figure}
\includegraphics[%
  width=10cm]{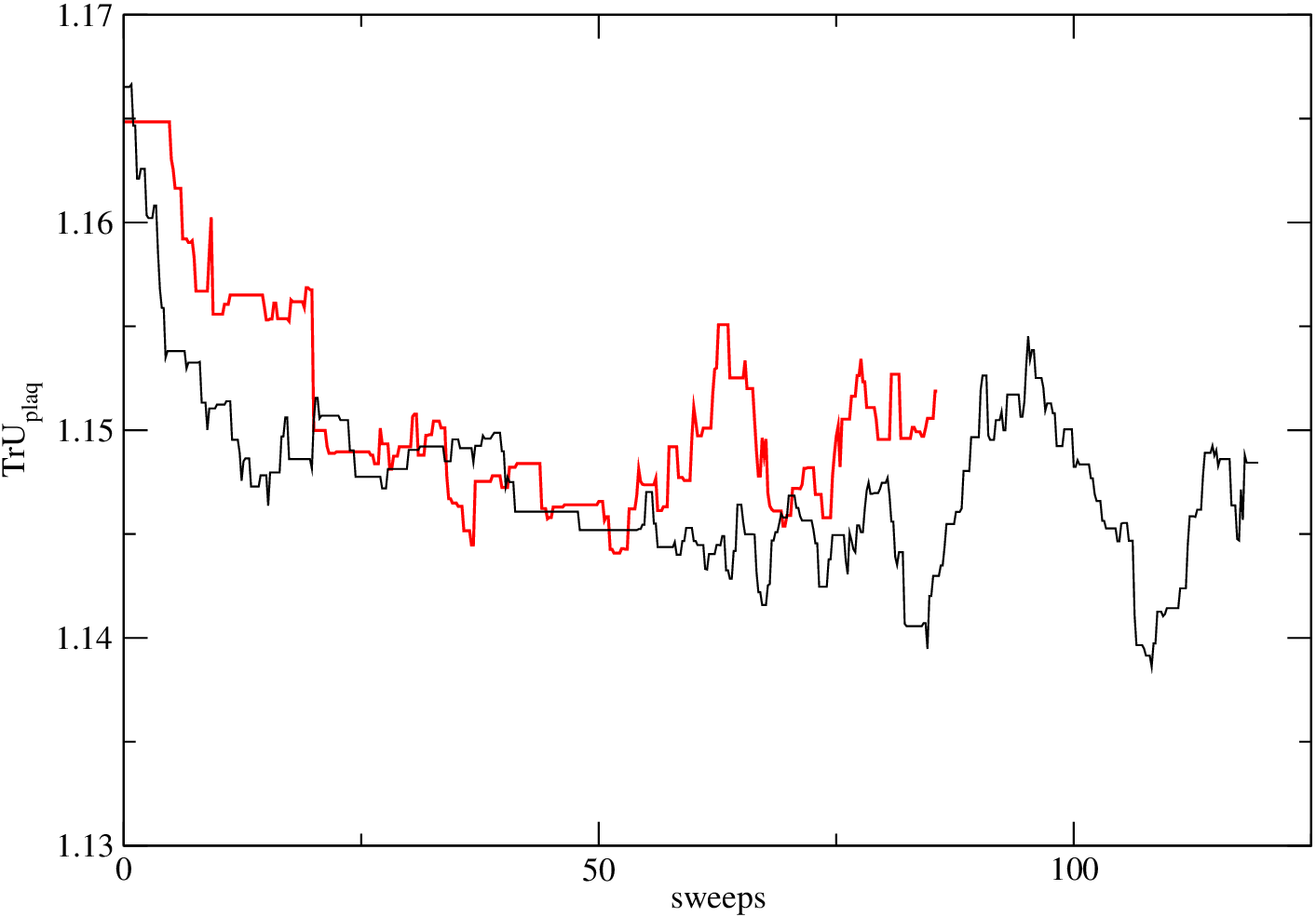}

\caption{The equilibration of the plaquette as the function of updating sweeps
in a $2+1$ flavor dynamical simulation. The two series started from
two different quenched gauge configurations. One sweep on the horizontal
axis corresponds to a double update of the whole lattice.\label{Fig:plaq_vs_sweep}}
\end{figure}
Figure \ref{Fig:plaq_vs_sweep} shows the equilibration of the plaquette
as the function of updating steps for our two sets. The units on the
horizontal axis are given in sweeps that correspond to a double update
of the whole lattice which corresponds to about 5 full cycles. We
determined the lattice spacing from the static potential as $a=0.14(1)$fm,
which gives the spatial size $L_{s}\approx1.1$fm. On this rather
small volume the hadron spectrum shows large finite volume effects.
Instead of the pion mass, which is dominated by the volume, we estimate
the quark mass from the eigenvalue spectrum of the Dirac operator. 

\begin{figure}
\begin{center}\includegraphics[%
  width=10cm]{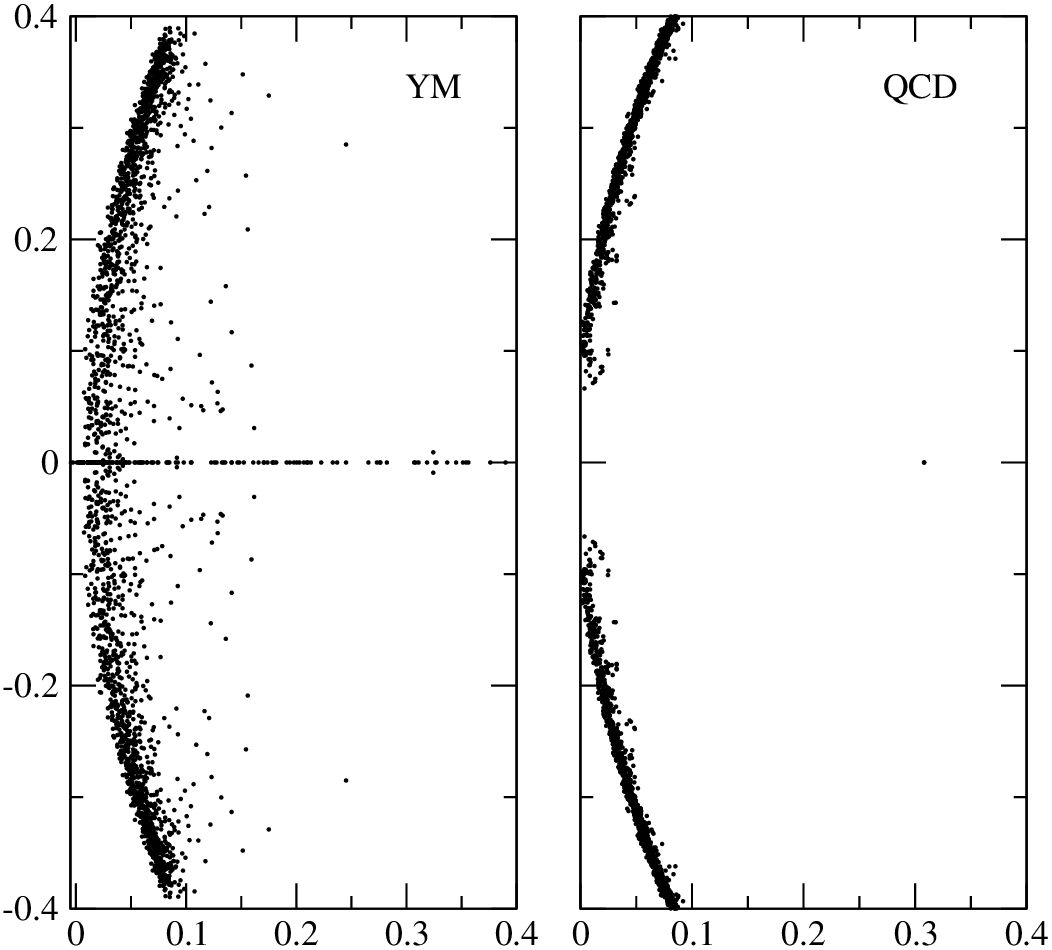}\end{center}

\caption{The low lying eigenvalue spectrum of the Dirac operator on 50 pure
Yang-Mills (left) and 50 equilibrated dynamical (right) configurations.
Both panels correspond to $8^{3}\times24$, $a\approx0.14\,$fm lattices
with the same lattice quark masses. \label{Fig:Eig_spect_low} }
\end{figure}

The left panel of Figure \ref{Fig:Eig_spect_low} shows the first
48 low energy eigenvalues on 50 $a\simeq0.15\,$fm pure Yang-Mills
configurations with $m_{q}=0.017$. This quark mass corresponds to
$m_{\pi}\simeq300\,$MeV pions in the quenched approximation. The
right panel shows the first 48 low lying eigenvalues on 50 equilibrated
dynamical configurations. The eigenvalues of a massless chiral Dirac
operator lie on circle if $R={\rm {const}}$. Our Dirac operator has
a non-trivial $R$ in which case one knows crude bounds only:the eigenvalues
should lie between two circles touching each other at the origin.
The full spectrum on a $4^{4}$configuration in Fig. \ref{Fig:spect_1}
gives more information. A small quark mass shifts the eigenvalues
to the right. The scattering of the eigenmodes characterize the chiral
symmetry breaking of our approximate Dirac operator. 

While the scatter of the eigenmodes on the left panel is not negligible,
its scale is small (compare to the whole spectrum of Fig. \ref{Fig:spect_1}).
In the quenched hadron spectrum calculations of 
Refs. \cite{Hasenfratz:2002rp,Hasenfratz:2004qk,Gattringer:2003qx}
at similar parameters no exceptional configurations were observed
which is consistent with the quenched spectrum in Fig. \ref{Fig:Eig_spect_low}
here. 

There are several new features we can identify on the right panel
that corresponds to the dynamical configurations. First we observe
that the eigenmodes are shifted somewhat to the left. Their value
suggests a nearly zero physical quark mass indicating a small additive
mass renormalization of $\delta m\approx -0.015$. On the quenched
configurations 
$\delta m$ is practically zero. As we based our parameters on the
quenched spectroscopy, by accident we simulated an approximately massless
dynamical quark  system. Since in the 2nd A/R step of the update (Sect.
\ref{sub:The-2nd-accept/reject}) we subtract the low lying eigenmodes,
this did not cause any increase in computing time. 

On its own a small mass renormalization is not a problem. It is the
fluctuations of the eigenmodes beyond $\delta m$ that create exceptional
configurations. These fluctuation are suppressed on the dynamical
configurations as compared to the quenched case. The two panels of
Figure \ref{Fig:Eig_spect_low} contain the same number of eigenvalues
and it is apparent that the Dirac operator on the dynamical configurations
is much more chiral than on the quenched ones. This unexpected benefit
is the effect of the fermionic determinant in the Boltzmann weight
and shows that its presence enhances the configurations where our
parametrization of the fixed point Dirac operator works better. 

The very small eigenmodes, those with $|\lambda|<0.1$, are completely
missing from the right panel. This is the consequence of the suppression
of the low eigenmodes by the determinant. The gauge update of Sect.\ref{sub:Gauge-update}
creates configurations with real eigenmodes and some of these are
accepted by the A/R steps. One of these modes is present on the right
panel of Figure \ref{Fig:Eig_spect_low} at $\lambda\approx0.3$.
However as these real eigenvalues move toward zero their determinants
become small and the configurations are eventually replaced by configurations
without real eigenmodes. On large volumes the small complex eigenmodes
are not completely suppressed but on these small volumes even those
are missing. 

\section{Conclusion}

In this paper we discussed the Partial Global Stochastic Update method
and its use to simulate dynamical fixed point fermions. The original
method is combined with several improvement techniques. By separating
and controlling both the UV and IR modes of the Dirac operator and
the fluctuations of its stochastic estimator we found the update efficient
on moderate volumes even near the chiral limit. To illustrate the
algorithm we presented some preliminary results on $8^{3}\times24,$
$a\approx0.14\,$fm lattices with 2+1 flavors with an approximately
massless light doublet.

\begin{acknowledgments}
We thank Stefan Solbrig for his help in the optimization of the code
on Hitachi SR8000. 
A.H. acknowledges support from the US Department of Energy.
This work was supported in part by Schweizerischer Nationalfonds.
\end{acknowledgments}
\bibliographystyle{apsrev}
\bibliography{lattice}

\begin{thebibliography}{30}
\expandafter\ifx\csname natexlab\endcsname\relax\def\natexlab#1{#1}\fi
\expandafter\ifx\csname bibnamefont\endcsname\relax
  \def\bibnamefont#1{#1}\fi
\expandafter\ifx\csname bibfnamefont\endcsname\relax
  \def\bibfnamefont#1{#1}\fi
\expandafter\ifx\csname citenamefont\endcsname\relax
  \def\citenamefont#1{#1}\fi
\expandafter\ifx\csname url\endcsname\relax
  \def\url#1{\texttt{#1}}\fi
\expandafter\ifx\csname urlprefix\endcsname\relax\def\urlprefix{URL }\fi
\providecommand{\bibinfo}[2]{#2}
\providecommand{\eprint}[2][]{\url{#2}}

\bibitem[{\citenamefont{Hasenfratz et~al.}(2001)}]{Hasenfratz:2000xz}
\bibinfo{author}{\bibfnamefont{P.}~\bibnamefont{Hasenfratz}}
  \bibnamefont{et~al.}, \bibinfo{journal}{Int. J. Mod. Phys.}
  \textbf{\bibinfo{volume}{C12}}, \bibinfo{pages}{691} (\bibinfo{year}{2001}),
  \eprint{hep-lat/0003013}.

\bibitem[{\citenamefont{Niedermayer et~al.}(2001)\citenamefont{Niedermayer,
  Rufenacht, and Wenger}}]{Niedermayer:2000yx}
\bibinfo{author}{\bibfnamefont{F.}~\bibnamefont{Niedermayer}},
  \bibinfo{author}{\bibfnamefont{P.}~\bibnamefont{Rufenacht}},
  \bibnamefont{and} \bibinfo{author}{\bibfnamefont{U.}~\bibnamefont{Wenger}},
  \bibinfo{journal}{Nucl. Phys.} \textbf{\bibinfo{volume}{B597}},
  \bibinfo{pages}{413} (\bibinfo{year}{2001}),
  \eprint[http://arXiv.org/abs]{hep-lat/0007007}.

\bibitem[{\citenamefont{Jorg}(2002)}]{Jorg:2002nm}
\bibinfo{author}{\bibfnamefont{T.}~\bibnamefont{Jorg}} (\bibinfo{year}{2002}),
  \eprint{hep-lat/0206025}.

\bibitem[{\citenamefont{Hauswirth}(2002)}]{Hauswirth:2002eq}
\bibinfo{author}{\bibfnamefont{S.}~\bibnamefont{Hauswirth}}
  (\bibinfo{year}{2002}), \eprint{hep-lat/0204015}.

\bibitem[{\citenamefont{Gattringer et~al.}(2004)}]{Gattringer:2003qx}
\bibinfo{author}{\bibfnamefont{C.}~\bibnamefont{Gattringer}}
  \bibnamefont{et~al.} (\bibinfo{collaboration}{BGR}), \bibinfo{journal}{Nucl.
  Phys.} \textbf{\bibinfo{volume}{B677}}, \bibinfo{pages}{3}
  (\bibinfo{year}{2004}), \eprint{hep-lat/0307013}.

\bibitem[{\citenamefont{Hasenfratz and Niedermayer}(1994)}]{Hasenfratz:1993sp}
\bibinfo{author}{\bibfnamefont{P.}~\bibnamefont{Hasenfratz}} \bibnamefont{and}
  \bibinfo{author}{\bibfnamefont{F.}~\bibnamefont{Niedermayer}},
  \bibinfo{journal}{Nucl. Phys.} \textbf{\bibinfo{volume}{B414}},
  \bibinfo{pages}{785} (\bibinfo{year}{1994}), \eprint{hep-lat/9308004}.

\bibitem[{\citenamefont{Ginsparg and Wilson}(1982)}]{Ginsparg:1981bj}
\bibinfo{author}{\bibfnamefont{P.~H.} \bibnamefont{Ginsparg}} \bibnamefont{and}
  \bibinfo{author}{\bibfnamefont{K.~G.} \bibnamefont{Wilson}},
  \bibinfo{journal}{Phys. Rev.} \textbf{\bibinfo{volume}{D25}},
  \bibinfo{pages}{2649} (\bibinfo{year}{1982}).

\bibitem[{\citenamefont{Hasenfratz}(1998)}]{Hasenfratz:1997ft}
\bibinfo{author}{\bibfnamefont{P.}~\bibnamefont{Hasenfratz}},
  \bibinfo{journal}{Nucl. Phys. Proc. Suppl.} \textbf{\bibinfo{volume}{63}},
  \bibinfo{pages}{53} (\bibinfo{year}{1998}), \eprint{hep-lat/9709110}.

\bibitem[{\citenamefont{Hasenfratz
  et~al.}(2002{\natexlab{a}})\citenamefont{Hasenfratz, Hauswirth, Holland,
  Jorg, and Niedermayer}}]{Hasenfratz:2001qp}
\bibinfo{author}{\bibfnamefont{P.}~\bibnamefont{Hasenfratz}},
  \bibinfo{author}{\bibfnamefont{S.}~\bibnamefont{Hauswirth}},
  \bibinfo{author}{\bibfnamefont{K.}~\bibnamefont{Holland}},
  \bibinfo{author}{\bibfnamefont{T.}~\bibnamefont{Jorg}}, \bibnamefont{and}
  \bibinfo{author}{\bibfnamefont{F.}~\bibnamefont{Niedermayer}},
  \bibinfo{journal}{Nucl. Phys. Proc. Suppl.} \textbf{\bibinfo{volume}{106}},
  \bibinfo{pages}{751} (\bibinfo{year}{2002}{\natexlab{a}}),
  \eprint{hep-lat/0109007}.

\bibitem[{\citenamefont{Hasenfratz
  et~al.}(2002{\natexlab{b}})\citenamefont{Hasenfratz, Hauswirth, Jorg,
  Niedermayer, and Holland}}]{Hasenfratz:2002rp}
\bibinfo{author}{\bibfnamefont{P.}~\bibnamefont{Hasenfratz}},
  \bibinfo{author}{\bibfnamefont{S.}~\bibnamefont{Hauswirth}},
  \bibinfo{author}{\bibfnamefont{T.}~\bibnamefont{Jorg}},
  \bibinfo{author}{\bibfnamefont{F.}~\bibnamefont{Niedermayer}},
  \bibnamefont{and} \bibinfo{author}{\bibfnamefont{K.}~\bibnamefont{Holland}},
  \bibinfo{journal}{Nucl. Phys.} \textbf{\bibinfo{volume}{B643}},
  \bibinfo{pages}{280} (\bibinfo{year}{2002}{\natexlab{b}}),
  \eprint{hep-lat/0205010}.

\bibitem[{\citenamefont{Hasenfratz et~al.}(2004)\citenamefont{Hasenfratz, Juge,
  and Niedermayer}}]{Hasenfratz:2004qk}
\bibinfo{author}{\bibfnamefont{P.}~\bibnamefont{Hasenfratz}},
  \bibinfo{author}{\bibfnamefont{K.~J.} \bibnamefont{Juge}}, \bibnamefont{and}
  \bibinfo{author}{\bibfnamefont{F.}~\bibnamefont{Niedermayer}}
  (\bibinfo{collaboration}{Bern-Graz-Regensburg}), \bibinfo{journal}{JHEP}
  \textbf{\bibinfo{volume}{12}}, \bibinfo{pages}{030} (\bibinfo{year}{2004}),
  \eprint{hep-lat/0411034}.

\bibitem[{\citenamefont{Hasenfratz
  et~al.}(2002{\natexlab{c}})\citenamefont{Hasenfratz, Hauswirth, Holland,
  Jorg, and Niedermayer}}]{Hasenfratz:2001hr}
\bibinfo{author}{\bibfnamefont{P.}~\bibnamefont{Hasenfratz}},
  \bibinfo{author}{\bibfnamefont{S.}~\bibnamefont{Hauswirth}},
  \bibinfo{author}{\bibfnamefont{K.}~\bibnamefont{Holland}},
  \bibinfo{author}{\bibfnamefont{T.}~\bibnamefont{Jorg}}, \bibnamefont{and}
  \bibinfo{author}{\bibfnamefont{F.}~\bibnamefont{Niedermayer}},
  \bibinfo{journal}{Nucl. Phys. Proc. Suppl.} \textbf{\bibinfo{volume}{106}},
  \bibinfo{pages}{799} (\bibinfo{year}{2002}{\natexlab{c}}),
  \eprint{hep-lat/0109004}.

\bibitem[{\citenamefont{DeGrand et~al.}(1998)\citenamefont{DeGrand, Hasenfratz,
  and Kovacs}}]{DeGrand:1997ss}
\bibinfo{author}{\bibfnamefont{T.}~\bibnamefont{DeGrand}},
  \bibinfo{author}{\bibfnamefont{A.}~\bibnamefont{Hasenfratz}},
  \bibnamefont{and} \bibinfo{author}{\bibfnamefont{T.~G.}
  \bibnamefont{Kovacs}}, \bibinfo{journal}{Nucl. Phys.}
  \textbf{\bibinfo{volume}{B520}}, \bibinfo{pages}{301} (\bibinfo{year}{1998}),
  \eprint{hep-lat/9711032}.

\bibitem[{\citenamefont{Lang et~al.}(2004)\citenamefont{Lang, Majumdar, and
  Ortner}}]{Lang:2004ze}
\bibinfo{author}{\bibfnamefont{C.~B.} \bibnamefont{Lang}},
  \bibinfo{author}{\bibfnamefont{P.}~\bibnamefont{Majumdar}}, \bibnamefont{and}
  \bibinfo{author}{\bibfnamefont{W.}~\bibnamefont{Ortner}}
  (\bibinfo{year}{2004}), \eprint{hep-lat/0412016}.

\bibitem[{\citenamefont{Hasenfratz and Knechtli}(2002)}]{Hasenfratz:2002jn}
\bibinfo{author}{\bibfnamefont{A.}~\bibnamefont{Hasenfratz}} \bibnamefont{and}
  \bibinfo{author}{\bibfnamefont{F.}~\bibnamefont{Knechtli}},
  \bibinfo{journal}{Comput. Phys. Commun.} \textbf{\bibinfo{volume}{148}},
  \bibinfo{pages}{81} (\bibinfo{year}{2002}),
  \eprint[http://arXiv.org/abs]{hep-lat/0203010}.

\bibitem[{\citenamefont{Grady}(1985)}]{Grady:1985fs}
\bibinfo{author}{\bibfnamefont{M.}~\bibnamefont{Grady}},
  \bibinfo{journal}{Phys. Rev.} \textbf{\bibinfo{volume}{D32}},
  \bibinfo{pages}{1496} (\bibinfo{year}{1985}).

\bibitem[{\citenamefont{Creutz}(1992)}]{Creutz:Book}
\bibinfo{author}{\bibfnamefont{M.}~\bibnamefont{Creutz}},
  \bibinfo{journal}{Algorithms for Simulating Fermions}
  (\bibinfo{year}{1992}), \bibinfo{note}{in Quantum Fields on the Computer,
  World Scientific Publishing}.

\bibitem[{\citenamefont{Hasenfratz and Alexandru}(2002)}]{Hasenfratz:2002ym}
\bibinfo{author}{\bibfnamefont{A.}~\bibnamefont{Hasenfratz}} \bibnamefont{and}
  \bibinfo{author}{\bibfnamefont{A.}~\bibnamefont{Alexandru}},
  \bibinfo{journal}{Phys. Rev.} \textbf{\bibinfo{volume}{D65}},
  \bibinfo{pages}{114506} (\bibinfo{year}{2002}), \eprint{hep-lat/0203026}.

\bibitem[{\citenamefont{Alexandru and Hasenfratz}(2002)}]{Alexandru:2002jr}
\bibinfo{author}{\bibfnamefont{A.}~\bibnamefont{Alexandru}} \bibnamefont{and}
  \bibinfo{author}{\bibfnamefont{A.}~\bibnamefont{Hasenfratz}},
  \bibinfo{journal}{Phys. Rev.} \textbf{\bibinfo{volume}{D66}},
  \bibinfo{pages}{094502} (\bibinfo{year}{2002}), \eprint{hep-lat/0207014}.

\bibitem[{\citenamefont{Knechtli and Wolff}(2003)}]{Knechtli:2003yt}
\bibinfo{author}{\bibfnamefont{F.}~\bibnamefont{Knechtli}} \bibnamefont{and}
  \bibinfo{author}{\bibfnamefont{U.}~\bibnamefont{Wolff}}
  (\bibinfo{collaboration}{Alpha}), \bibinfo{journal}{Nucl. Phys.}
  \textbf{\bibinfo{volume}{B663}}, \bibinfo{pages}{3} (\bibinfo{year}{2003}),
  \eprint{hep-lat/0303001}.

\bibitem[{\citenamefont{Kennedy and Kuti}(1985)}]{Kennedy:1985pg}
\bibinfo{author}{\bibfnamefont{A.~D.} \bibnamefont{Kennedy}} \bibnamefont{and}
  \bibinfo{author}{\bibfnamefont{J.}~\bibnamefont{Kuti}},
  \bibinfo{journal}{Phys. Rev. Lett.} \textbf{\bibinfo{volume}{54}},
  \bibinfo{pages}{2473} (\bibinfo{year}{1985}).

\bibitem[{\citenamefont{Kennedy et~al.}(1988)\citenamefont{Kennedy, Kuti,
  Meyer, and Pendleton}}]{Kennedy:1988yy}
\bibinfo{author}{\bibfnamefont{A.~D.} \bibnamefont{Kennedy}},
  \bibinfo{author}{\bibfnamefont{J.}~\bibnamefont{Kuti}},
  \bibinfo{author}{\bibfnamefont{S.}~\bibnamefont{Meyer}}, \bibnamefont{and}
  \bibinfo{author}{\bibfnamefont{B.~J.} \bibnamefont{Pendleton}},
  \bibinfo{journal}{Phys. Rev.} \textbf{\bibinfo{volume}{D38}},
  \bibinfo{pages}{627} (\bibinfo{year}{1988}).

\bibitem[{\citenamefont{Joo et~al.}(2003)\citenamefont{Joo, Horvath, and
  Liu}}]{Joo:2001bz}
\bibinfo{author}{\bibfnamefont{B.}~\bibnamefont{Joo}},
  \bibinfo{author}{\bibfnamefont{I.}~\bibnamefont{Horvath}}, \bibnamefont{and}
  \bibinfo{author}{\bibfnamefont{K.~F.} \bibnamefont{Liu}},
  \bibinfo{journal}{Phys. Rev.} \textbf{\bibinfo{volume}{D67}},
  \bibinfo{pages}{074505} (\bibinfo{year}{2003}), \eprint{hep-lat/0112033}.

\bibitem[{\citenamefont{Montvay and Scholz}(2005)}]{Montvay:2005tj}
\bibinfo{author}{\bibfnamefont{I.}~\bibnamefont{Montvay}} \bibnamefont{and}
  \bibinfo{author}{\bibfnamefont{E.}~\bibnamefont{Scholz}}
  (\bibinfo{year}{2005}), \eprint{hep-lat/0506006}.

\bibitem[{\citenamefont{Hasenbusch}(2001)}]{Hasenbusch:2001ne}
\bibinfo{author}{\bibfnamefont{M.}~\bibnamefont{Hasenbusch}},
  \bibinfo{journal}{Phys. Lett.} \textbf{\bibinfo{volume}{B519}},
  \bibinfo{pages}{177} (\bibinfo{year}{2001}),
  \eprint[http://arXiv.org/abs]{hep-lat/0107019}.

\bibitem[{\citenamefont{de~Forcrand}(1999)}]{deForcrand:1998sv}
\bibinfo{author}{\bibfnamefont{P.}~\bibnamefont{de~Forcrand}},
  \bibinfo{journal}{Nucl. Phys. Proc. Suppl.} \textbf{\bibinfo{volume}{73}},
  \bibinfo{pages}{822} (\bibinfo{year}{1999}), \eprint{hep-lat/9809145}.

\bibitem[{\citenamefont{Alexandru and Hasenfratz}(2003)}]{Alexandru:2002sw}
\bibinfo{author}{\bibfnamefont{A.}~\bibnamefont{Alexandru}} \bibnamefont{and}
  \bibinfo{author}{\bibfnamefont{A.}~\bibnamefont{Hasenfratz}},
  \bibinfo{journal}{Nucl. Phys. Proc. Suppl.} \textbf{\bibinfo{volume}{119}},
  \bibinfo{pages}{997} (\bibinfo{year}{2003}), \eprint{hep-lat/0209070}.

\bibitem[{\citenamefont{Hasenbusch}(1999)}]{Hasenbusch:1998yb}
\bibinfo{author}{\bibfnamefont{M.}~\bibnamefont{Hasenbusch}},
  \bibinfo{journal}{Phys. Rev.} \textbf{\bibinfo{volume}{D59}},
  \bibinfo{pages}{054505} (\bibinfo{year}{1999}), \eprint{hep-lat/9807031}.

\bibitem[{\citenamefont{Montvay}(1998)}]{Montvay:1997vh}
\bibinfo{author}{\bibfnamefont{I.}~\bibnamefont{Montvay}},
  \bibinfo{journal}{Comput. Phys. Commun.} \textbf{\bibinfo{volume}{109}},
  \bibinfo{pages}{144} (\bibinfo{year}{1998}), \eprint{hep-lat/9707005}.

\bibitem[{\citenamefont{Borici and de~Forcrand}(1995)}]{Borici:1995am}
\bibinfo{author}{\bibfnamefont{A.}~\bibnamefont{Borici}} \bibnamefont{and}
  \bibinfo{author}{\bibfnamefont{P.}~\bibnamefont{de~Forcrand}},
  \bibinfo{journal}{Nucl. Phys.} \textbf{\bibinfo{volume}{B454}},
  \bibinfo{pages}{645} (\bibinfo{year}{1995}), \eprint{hep-lat/9505021}.

\end{thebibliography}

\end{document}